\documentclass[preprint,a4paper,showpacs,aps,prc]{revtex4-1}
\usepackage{graphicx}

\begin{document}

\title{First result of the experimental search for the
$2K$-capture of $^{124}$Xe with the copper proportional
counter\footnote{{\small Talk at the International Workshop
on Prospects of Particle Physics: "Neutrino Physics and
Astrophysics" Jan 26 - Ferb 2, 2014, Valday, Russia.}}} 

\author{
Yu.M.~Gavrilyuk$^\dag$, A.M.~Gangapshev$^\dag$, V.V.~Kazalov$^\dag$, V.V.~Kuzminov$^\dag$, 
S.I.~Panasenko$^\ddag$,
S.S.~Ratkevich$^\ddag$, D.A.~Tekueva$^\dag$ and S.P.~Yakimenko$^\dag$}

\affiliation{
$^\dag$Baksan Neutrino Observatory INR RAS, Russiaa \\
$^\ddag$ V.N.Karazin Kharkiv National University,Ukraine}%

\date{\today}%

\begin{abstract}
First result of experiment for searching of $2K$-capture of
$^{124}$Xe with the large-volume copper proportional counter (LPC)
is given. The 12 $L$ sample with 63.3\% (44 g) of $^{124}$Xe was
used in measurements. The limit on the half-life of $^{124}$Xe
with regard to $2K(2\nu)$-capture for the ground state of
$^{124}$Te ($0^+\rightarrow0^+$, g.s.) has been found:
$T_{1/2}\geq 4.67\times10^{20}$ y (90\% C.L.). A sample with volume 52 $L$
comprising of $^{124}$Xe (10.6 $L$ -58.6 g) and $^{126}$Xe (14.1 $L$ - 79.3~g)
will used at the next step of the experiment to increase
a sensitivity of 2K-caption of $^{124}$Xe registration.
In this case sensitivity to the investigated process will be at the level of $S =
1.46\times10^{21}$ y (90\% C.L.) for 1 year measurement.
\end{abstract}


\maketitle

\section{\label{sec:intro}Introduction}

The Baksan Neutrino Observatory INR RAS has several unique low
background laboratories. One of them is deep underground
low-background laboratory (DULB-4900) \cite{r1}. The laboratory is
located at a distance of 3700 m from the main entry to the Baksan
Neutrino Observatory tunnel, in the hall of $\sim
6\times6\times40$ m${^3}$. Thickness of the mountain rock over
DULB corresponds to 4900 m.w.e., thereby decreasing cosmic ray
flux by $\sim 10{^7}$ times. These allow one to carry out
of experiments for searching of rare process and decay.
An experiment for searching of $2K$-capture of $^{124}$Xe is an example.

Theoretical calculations for $2K(2\nu)$ within the frames of
different models (QRPA, MCM) predict the following half-life
periods with respect to $2K$-capture of $^{124}$Xe:
$1.08\times10^{22}$ y \cite{r2}, $3.9\times10^{23}$ y \cite{r3}. The
values were obtained by taking the fraction of $2K(2\nu)$ capture
events in $^{124}$Xe with respect to the total number of
$2e(2\nu)$ capture events to be 73\% \cite{r4}.

\section{Basic assumptions}

When two electrons are captured from the \emph{K}-shell in
$^{124}$Xe, a daughter atom of $^{124}$Te$^{**}$ is formed with
two vacancies in the \emph{K}-shell. The technique to search for
this reaction is based on the assumption that the energies of
characteristic photons and the probability that they will be
emitted when a double vacancy is filled are the same as the sum of
respective values when two single vacancies of the \emph{K}-shell
in two singly ionized Te$^*$ atoms are filled. In such a case, the
total measured energy is $2K_{ab}= 63.62$ keV, where ${K_{ab}}$ is
the binding energy of a $K$ electron in a Te atom (27.47  keV).
The fluorescence yield upon filling of a single vacancy in the
\emph{K}-shell of Te is 0.857. The energies and relative
intensities of the characteristic lines in the $K$ series are
$K_{\alpha 1}=27.47$~keV (100\%), $K_{\alpha 2}=27.20$~keV (54\%), $K_{\beta 1}=30.99$ keV
(18\%), and $K_{\beta 2}=31.7$ keV (5\%) \cite{r5}. There are three
possible ways for de-excitation of a doubly ionized
\emph{K}-shell: 1) emission of Auger electrons only $(e_a, e_a)$;
2) emission of a single characteristic quantum and an Auger
electron $(K,e_a)$; and 3) emission of two characteristic quanta
and low-energy Auger electrons $(K,K,e_a)$, with probabilities of
$p_1 = 0.020$, $p_2 = 0.246$, and $p_3 = 0.734$, respectively. A
characteristic quantum can travel a long enough distance in a gas
medium between the points of its production and absorption. But
photoelectrons produce almost pointwise charge clusters of primary
ionization in the gas. In case of the event with the escape of two
characteristic quanta absorbed in the working gas and a single
Auger electron, the energy is distributed among three pointwise
charge clusters. It is these three-point (or three-cluster) events
possessing a unique set of features that were the subject of the
search in Ref.\cite{r6}.

To register the process of $2K$-capture in $^{124}$Xe a large
proportional counter (LPC) with a casing of M1-grade copper has
been used \cite{PTE}. The LPC is inside the shielding of 18 cm
thick copper, 15 cm thick lead, and 8 cm thick borated
polyethylene layers. The installation is placed in one of the
chambers of the underground laboratory DULB-4900, where cosmic ray
flux is lowered to the level of $(3.03 \pm 0.10) \times 10^{-9}$ cm$^{-2}$s$^{-1}$ \cite{r8}.

The details of the counter signals formation, of the features of pulses
registration by digital oscilloscope, of the data treatment and
analysis are described in Refs.\cite{PTE, Kr78}.

\section{Measurement results}
The measurements with radioactive-pure xenon (background sample)
have been done at the initial stage of the experiment for understanding
background level of experimental setup and for future comprising
with $^{124}$Xe. The LPC pressure is equal 1.9 atm.
The measurement time was 970 hours. Count rate in
interval $15\div150$~keV was found to be $\sim 16$ h$^{-1}$.

A sample of 12~$L$  of xenon enriched in the isotope
$^{124}$Xe to 63.3\% (44~g) was used in the main measurements with $^{124}$Xe.
This sample is combined from several one, see Tab.\ref{t1}.
\begin{table*}[pt]
\caption{\label{t1} Characteristics of samples 124ех.}

\begin{tabular}
{|l| c c c c c c c c c|} \hline \hline

 \multicolumn{1}{|c} ~ &  \multicolumn{9}{|c|}{Xenon isotopes}   \\
 \cline{2-10}
 \multicolumn{1}{|c|} {Samples, \emph{L}} & {\small{124}} & {\small{126}} & {\small{128}} & {\small{129}} & {\small{130}} & {\small{131}} & {\small{132}}  & {\small{134}}  & {\small{136}} \\
  ~  &   \multicolumn{9}{c|}{Content, \emph{L} }  \\
\hline

$N^{\underline{o}}$~1a, {\small{1.98}}& {\small{1.02}} & {\small{3.81$\cdot10^{-1}$}} & {\small{5.19$\cdot10^{-1}$}}  &  {\small{3.1$\cdot10^{-4}$}} & {\small{1.3$\cdot10^{-4}$}} & {\small{3.76$\cdot10^{-5}$}} & {\small{1.8$\cdot10^{-5}$}} & {\small{1.8$\cdot10^{-5}$}} & {\small{1.8e$\cdot10^{-5}$}}\\

$N^{\underline{o}}$~2, {\small{1.27}} & {\small{1.13}} & {\small{1.3$\cdot10^{-2}$}} & {\small{3.4$\cdot10^{-2}$}}  &  {\small{6.1$\cdot10^{-4}$}} & {\small{1.2$\cdot10^{-4}$}} & {\small{3.6$\cdot10^{-5}$}} & {\small{3.6$\cdot10^{-5}$}} & {\small{3.5$\cdot10^{-5}$}} & {\small{3.5$\cdot10^{-5}$}}\\

$N^{\underline{o}}$~3, {\small{3.56}} & {\small{1.60}} & {\small{1.93}} & {\small{1.03$\cdot10^{-3}$}}  &  {\small{6.8$\cdot10^{-5}$}} & {\small{6.8$\cdot10^{-5}$}} & {\small{6.7$\cdot10^{-5}$}} & {\small{6.7$\cdot10^{-5}$}} & {\small{6.6$\cdot10^{-5}$}} & {\small{6.5$\cdot10^{-5}$}}\\

$N^{\underline{o}}$~4, {\small{3.11}} & {\small{1.50}} & {\small{1.4}} & {\small{1.2$\cdot10^{-1}$}}  &  {\small{6.7$\cdot10^{-4}$}} & {\small{6.7$\cdot10^{-4}$}} & {\small{5.8$\cdot10^{-4}$}} & {\small{5.8$\cdot10^{-4}$}} & {\small{5.8$\cdot10^{-4}$}} & {\small{5.8$\cdot10^{-4}$}}\\

$N^{\underline{o}}$~5, {\small{1.32}} & {\small{1.11}} & {\small{4.0$\cdot10^{-3}$}} & {\small{7.5$\cdot10^{-5}$}}  &  {\small{5.0$\cdot10^{-5}$}} & {\small{5.0$\cdot10^{-5}$}} & {\small{3.1$\cdot10^{-4}$}} & {\small{5.0$\cdot10^{-5}$}} & {\small{5.0$\cdot10^{-5}$}} & {\small{5.0$\cdot10^{-5}$}}\\

$N^{\underline{o}}$~6, 1.31 & {\small{1.31}} & {\small{1.6$\cdot10^{-4}$}}& {\small{1.3$\cdot10^{-5}$}}  &  {\small{1.3$\cdot10^{-5}$}} & {\small{1.3$\cdot10^{-5}$}} & {\small{1.3$\cdot10^{-5}$}} & {\small{1.2$\cdot10^{-5}$}} & {\small{1.2$\cdot10^{-5}$}} & {\small{1.2$\cdot10^{-5}$}}\\

\hline \hline
\end{tabular}
\end{table*}

The LPC was filled to the maximum pressure $P_{\rm{max}}=1.1$~atm and
the first test measurement was done  during 310 hours.
Count rate in interval $15\div150$~keV was found to be $\sim 60$ h$^{-1}$.
Excess background was created by $^{85}$Kr decays.
This isotope is a part of natural krypton ($\sim 50$ ppm) which dissolved in xenon.

To reduce a background of krypton an attempt was made to "wash out"
of $^{124}$Xe sample by "clean krypton" without $^{85}$Kr.
This procedure is based on the assumption that xenon
will settle on a bottle wall at the liquid nitrogen
temperature ($-190^\circ$C) (the pressure of saturated vapor $1.93\times10^{-3}$ mmHg)
but krypton will stay in gas phase because it has a relatively
large value of saturated vapor pressure ($\sim1.8$ mmHg).
It was planned preliminary to add into the xenon a few
cm$^{3}$ of the krypton and than to repeat this procedure several times.
The rate of xenon treating was controlled by residual
background of LPC. Unfortunately this procedure did not
give any positive results. One can assume that
all krypton has been absorbed by xenon similar to a charcoal trap.

\begin{figure*}[pt]
\begin{center}
\includegraphics*[width=3.5in,angle=0.]{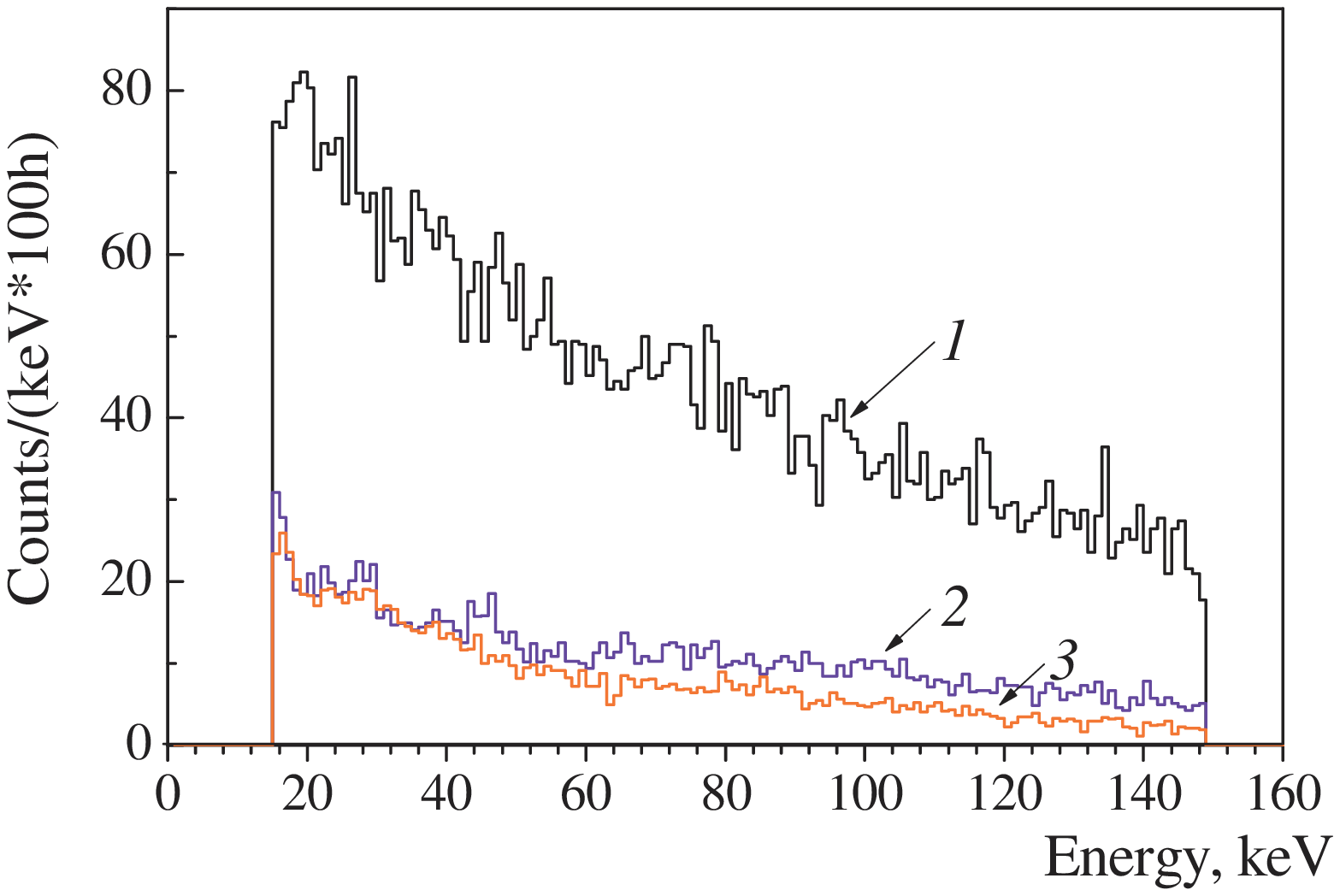}%
\caption{\label{f1}
The spectra of xenon before purification (\emph{1} -
$^{124}$Xe+$^{85}$Kr sample), radioactive-pure xenon (\emph{2} -
$^{nat}$Xe sample) and xenon after purification (\emph{3} -
$^{124}$Xe sample), normalized to 100 h.}
\end{center}
\end{figure*}

A helium was chosen as an another "washing" gas.
The 0.5~$L$ of a helium was added to the bottle with the xenon per one procedure.
The bottle has been frozen and helium was exhausted. The background of LPC in interval
$15\div150$~keV became $\sim 12$ h$^{-1}$, i.e., the desired result was
achieved after six procedures. Easy and chip procedure for gas cleaning from remnant
of krypton was developed as a result of this work.
A comparison of spectra of $^{124}$Xe sample before purification (spectrum \emph{1}),
of the radioactive-pure xenon (spectrum \emph{2}) and of
$^{124}$Xe after purification - (spectrum \emph{3}) are shown on Fig.\ref{f1}.
All spectra are normalized to 100 hours.

\begin{figure*} [pt]
\begin{center}
\includegraphics*[width=3.5in,angle=270.]{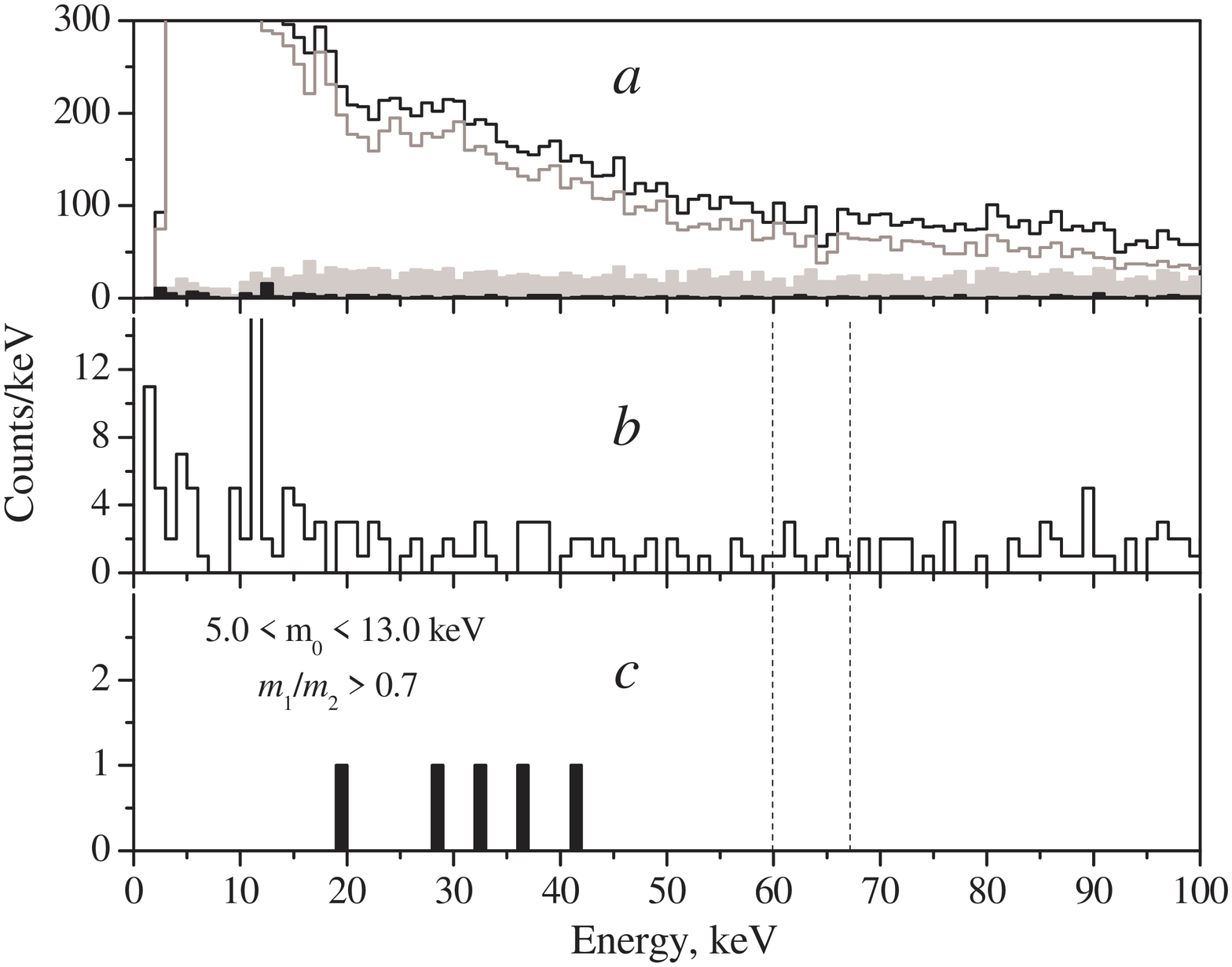}%
\caption{\label{f3} Amplitude spectra of LPC, 1.1 atm of xenon
enriched in the isotope $^{124}$Xe to 63.3\% (44 g): \emph{a}) the black
line - all events; the gray line - one-point events; the gray bar
graph - two-point events; the black bar graph - three-point
events; \emph{b}) the three-point event spectra - full; \emph{c}) the
three-point event spectra - selected by criteria ($5.0 \leq m_0
\leq 13.0$) keV and $m_1/m_2 \leq 0.7$. The pure measurement time
is 1130 h. The dashed lines indicate the boundaries of the region
of interest - ($63.6 \pm 3.7$) keV.}
\end{center}
\end{figure*}

The LPC was filled to the maximum possible for xenon sample pressure - 1.1 atm
to perform the main measurements for search of $2K$-capture of $^{124}$Xe.
The efficiency of registration of two $K$-photons
with energy $\sim27$~keV is equal 0.09 at this pressure.
Measurements in underground low-background setup were
carried out during 1130 h. The charge pulses are recorded by
digital oscilloscope to the personal computer. The collected data
set was treated by special program. As a result, the entire
dataset was divided into groups (spectra): one-, two- and
three-point events. The full spectrum, the spectra of one-, two-
and three-point events are shown on Fig.\ref{f3}$a$
by the black
line, the gray line, the gray bar graph and the black bar graph.
The three-point events spectrum is shown on Fig.\ref{f3}$b$ more precisely.
The "useful" effect is searching in three-point event
spectrum in energy interval $(63.6\pm3.7)$ keV.
The amplitudes of partial pulses for each event are arranged in the increased order
$ [(A_1,A_2,A_3)\rightarrow(m_0\leq m_1\leq m_2)]$, to simplify the selection
of events with a given set of features. Selected useful events must satisfy
the selection criteria ($5.0 \leq m_0\leq 13.0$) keV and $m_1/m_2\ge 0.7$.
The events corresponding to these ratios were selected from the spectrum
of the three-point pulses to reduce the background.
They are presented in Fig.\ref{f3}$c$.

The area of the expected effect marked by dashed lines.
There are not remained any events in this area after the event selection.
Therefore, we obtain $N_{eff} = 2.44$ for 1130 h $(\mu=0.00\div2.44)$
or $N_{eff} = 21.37$ y$^{-1}$ (90\% C.L.) by using of the recommendation of
the work \cite{r10} for the given (effect+background) value (0 events).
The half-life period has been calculated using formula:
\begin{eqnarray*}
\mathrm{lim}
\begin{array}{*{3}c}
   {} \\
\end{array}
T_{1/2}  \geq \mathrm{ln}2 \times N  \times \frac{p_3 \times
\varepsilon_p \times \varepsilon_3 } {N_{eff}},
\end{eqnarray*}
where $N=2.59\times10^{23}$ is the number of $^{124}$Xe atoms in
the operating volume of the counter, $p_3=0.735$ is a portion of
$2K$-captures accompanied by the emission of two-quanta;
$\varepsilon_p=0.09$ is the probability of two \emph{K}-quanta
absorption in the operating volume; $\varepsilon_3=0.422$ is the
selection efficiency for three-point events due to $2K$-capture in
$^{124}$Xe.

The result obtained is:
\begin{eqnarray*}
    T_{1/2} (0\nu+2\nu,2K) \geq 4.6\times 10^{20}\texttt{y}~(90\%~\texttt{C.L.}).
\end{eqnarray*}

It should be noted that for the "normal" working pressure 5 atm
efficiency of registration of two $K$-quanta is equal 0.809. A
sensitivity of experimental setup to the required process at this
value of efficiency is equal:
\begin{eqnarray*}
    S  = 1.46 \times 10^{21}\texttt{y}~(90\%~\texttt{C.L.}).
\end{eqnarray*}

\begin{table*}
\caption{\label{t2} Characteristics of samples
$N^{\underline{o}}$~7, total vol. $\sim$ 58 \emph{L}, $^{124}$Xe
$\sim$ (4.33~\emph{L}~$\approx$~23.96~g), $^{126}$Xe $\sim$
(15.24~\emph{L} $\approx$~85.7~g).}

\begin{tabular}
{|l| c c c c c c c c c|} \hline \hline
 \multicolumn{1}{|c} ~ &  \multicolumn{9}{|c|}{Xenon isotopes}   \\
 \cline{2-10}
 \multicolumn{1}{|c|} {Samples, \emph{L}} & {\small{124}} & {\small{126}} & {\small{128}} & {\small{129}} & {\small{130}} & {\small{131}} & {\small{132}}  & {\small{134}}  & {\small{136}} \\
  ~  &   \multicolumn{9}{c|}{Content, \emph{L} }  \\
\hline

$N^{\underline{o}}$~7, {\small{58}} & {\small{4.33}} & {\small{15.24}} & {\small{24.147}}  & {\small{14.033}} &  {\small{0.0511}} & {\small{0.0394}} & {\small{0.0168}} & {\small{0.0606}} & {\small{0.0543}}\\

\hline \hline
\end{tabular}
\end{table*}

A new 58~$L$ sample $N^{\underline{o}}$~7 of the enriched xenon was
obtained by the team at the last time. The characteristics of the sample are shown in
Tab.\ref{t2}.

\begin{figure*} [ht]
\begin{center}
\includegraphics*[width=3.25in,angle=0.]{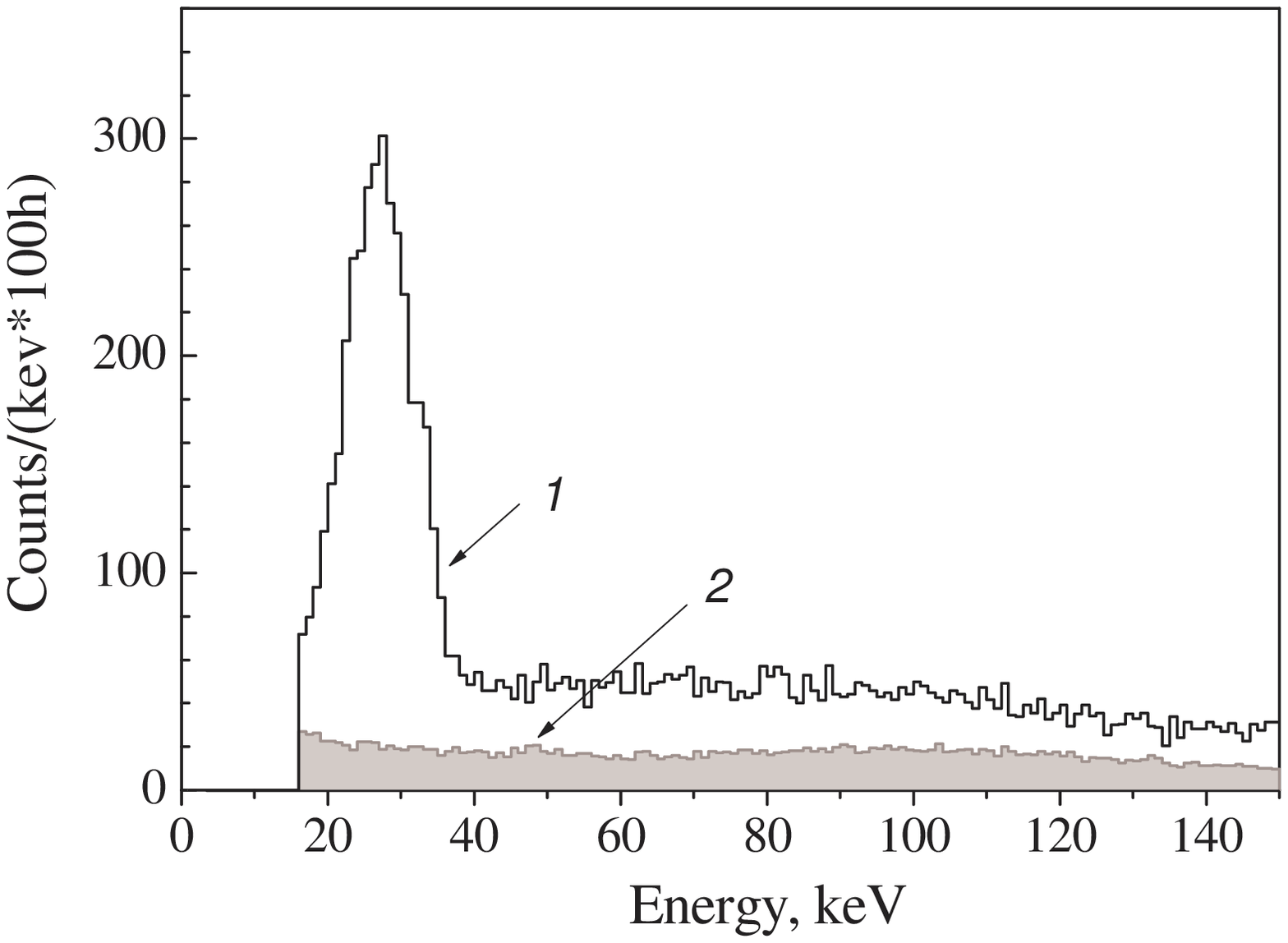}%
\caption{\label{f4} Amplitude spectra of LPC with xenon sample
$N^{\underline{o}}$~7 (5 atm) - spectrum \emph{1} and
radioactive-pure xenon (5 atm) - spectrum \emph{2}.}
\end{center}
\end{figure*}
\begin{figure*}[pt]
\begin{center}
\includegraphics*[width=3.50in,angle=0.]{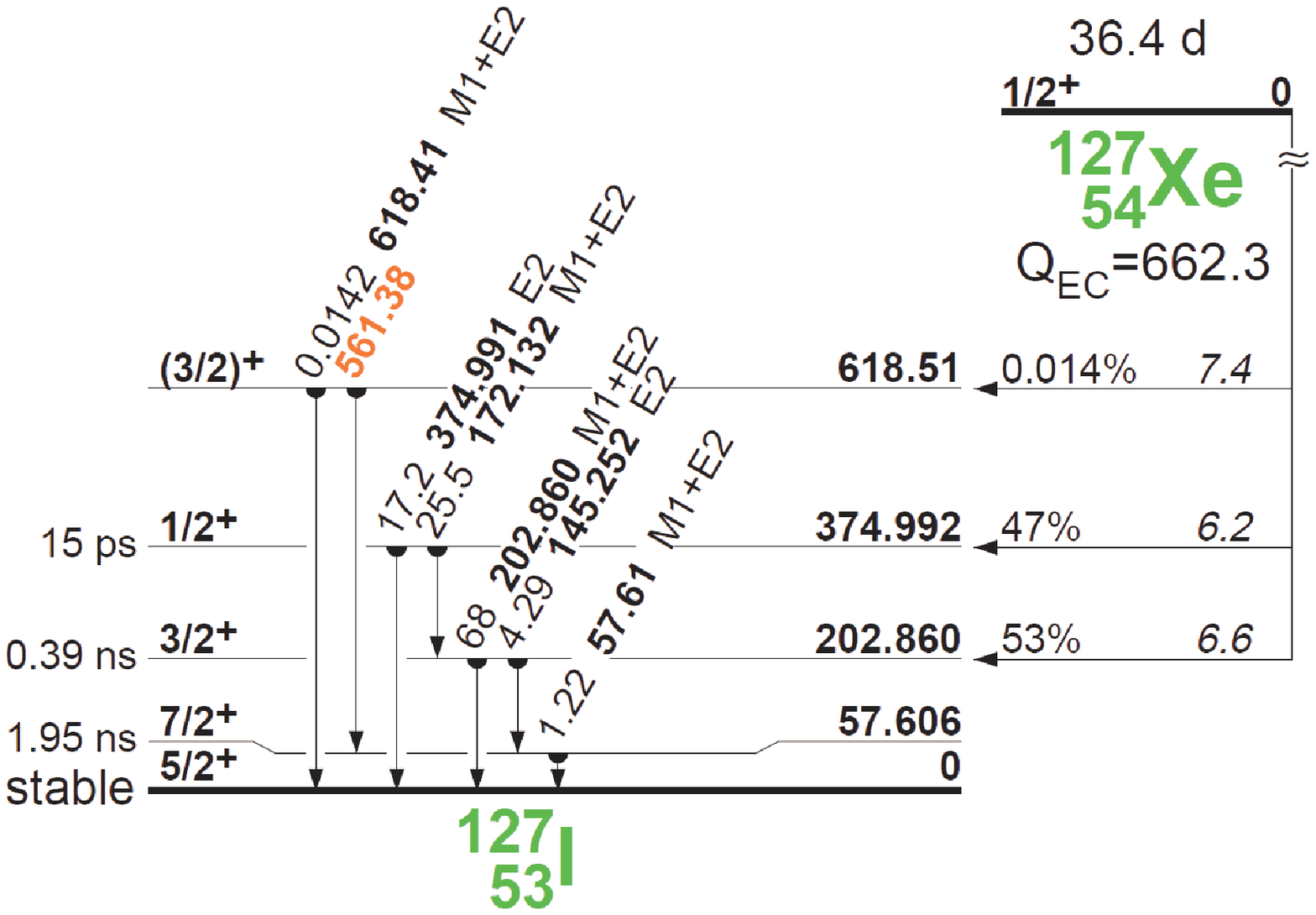}%
\caption{\label{f5} The decay scheme of isotope $^{127}$Xe
\cite{r11}. }
\end{center}
\end{figure*}

The LPC was filled to 5 atm by new xenon sample
$N^{\underline{o}}$~7 for checking background measurement.
The spectrum (\emph{1}) of the sample $N^{\underline{o}}$~7 background
obtained at 146~h is shown on Fig.\ref{f4}.
The spectrum (\emph{2}) of the
radioactive-pure xenon is shown for comparison also.
All spectra are normalized to 100~h.
It is seen from the spectra comparison that the sample $N^{\underline{o}}$~7
has radioactive impurities. It creates a background in a wide energy range
including to the region of interest. Intense peak is visible in area of 28 keV.
Differences of the spectra (\emph{1}-\emph{2}) gives a count of rate $\sim
48$ h$^{-1}$ in the interval $15\div150$~keV. It turned out that this value decreases
exponentially over time with a half-life of about 30 days.
The radioactive isotope $^{127}$Xe was identified as the source due to the set of the attributes.
The decay scheme of isotope $^{127}$Xe \cite{r11} is shown in Fig.\ref{f5}.
The isotope $^{127}$Xe can be produce by secondary cosmic rays in isotopes of sample
$N^{\underline{o}}$~7 through reaction $^{126}$Xe$(n,\gamma)^{127}$Xe and $^{128}$Xe$(\mu,n)^{127}$Xe.
There are no published data on the cross sections of the reactions
mentioned above. Apparently, this can be explained by the fact
that in a natural xenon content of the original isotopes is
extremely small - $^{126}$Xe - 0.09\%, $^{128}$Xe - 1.91\%. Sample
$N^{\underline{o}}$~7 is highly enriched in these isotopes, and it
can be used to measure the previously mentioned cross sections.

The time of about 4 months is required for decrease of $^{127}$Xe background to
acceptable level before using of xenon sample $N^{\underline{o}}$~7 in the
experiment for searching of $2K$-capture of $^{124}$Xe.
A significant amount of $^{126}$Xe is present in the sample $N^{\underline{o}}$~7.
This isotope can also decay by $2K$ capture. The transition energy is 897 keV.
A high limit for the $2K$-capture half-life of $^{124}$Xe could be
obtained simultaneously if the sample $N^{\underline{o}}$~7 will be used in
experiment for searching of the $2K$-capture of $^{124}$Xe.

\section{Conclusions}

The new experiment for searching of $2K$-capture of $^{124}$Xe is done.
The 12~$L$ sample with 63.3\% (44~g) of $^{124}$Xe was
used in measurements. The limit on the half-life of $^{124}$Xe
with regard to $2K(2\nu)$-capture for the ground state of
$^{124}$Te ($0^+\rightarrow0^+$, g.s.) has been found:
$T_{1/2}\geq 4.67\times10^{20}$~y (90\% C.L.).

A new 52~$L$ sample of enriched xenon will used on the next stage of the experiment.
This sample has as part of $^{124}$Xe (10.6~$L$ -58.6 g) and $^{126}$Xe (14.1~$L$ - 79.3~g).
A sensitivity at 1 year measurement time of the experimental setup to the required process will be equal to
\begin{eqnarray*}
    S  = 1.46 \times 10^{21}\texttt{y}~(90\%~\texttt{C.L.}).
\end{eqnarray*}

\textbf{Acknowledgement.} The authors gratefully thank to the E.Yu.~Povolotskii for the advice on the purification of noble gases.

\end{document}